\documentclass[journal]{IEEEtran}

\usepackage{graphicx}
\usepackage{placeins}
\usepackage{amsmath}
\usepackage[linesnumbered,ruled]{algorithm2e}

\ifCLASSINFOpdf
\else
\fi

\begin{document}
%
\title{A Multi-Bit Neuromorphic Weight Cell using Ferroelectric FETs,
suitable for SoC Integration}
%
%
%

\author{Borna~Obradovic,
        Titash~Rakshit,
				Ryan~Hatcher,
				Jorge~Kittl,
				Rwik~Sengupta,
				Joon~Goo~Hong,
        and~Mark~S.~Rodder
\thanks{B. Obradovic, T. Rakshit, R. Hatcher, J. Kitttl, R. Sengupta, J.G. Hong and M. S. Rodder are with the 
Samsung Advanced Logic Lab, Austin TX.}
}
\maketitle

\begin{abstract}
A multi-bit digital weight cell for high-performance, inference-only non-GPU-like neuromorphic
accelerators is presented. The cell is designed with simplicity of peripheral circuitry
in mind. Non-volatile storage of weights which eliminates the need for DRAM access 
is based on FeFETs and is purely digital.
The Multiply-and-Accumulate operation is performed using passive resistors, gated by FeFETs.
The resulting weight cell offers a high degree of linearity and a large ON/OFF ratio.
The key performance tradeoffs are investigated, and the device requirements are elucidated.

\end{abstract}


%
\IEEEpeerreviewmaketitle

\section{Introduction}
%
%
%
%
\IEEEPARstart{H}{ardware} accelerators for Deep Neural Nets (DNNs) are receiving increased attention as Neural Network-based
applications proliferate. Currently, hardware accelerators fall into two categories: GPU-like devices \cite{GoogleTPU} with reduced-precision
arithmetic (and possibly an improved memory access architecture), and more "literal" implementations of DNNs as
sets of resistive cross-bar arrays (\cite{SYu1, SYu2, SYu3, GBurr1, GBurr2}). While the former approach is already seeing practical use, the latter is still in
the research phase. The key characteristics of the "literal" approach are local storage of network weights and analog computation
of the outputs as linear combinations of inputs. By storing network weights locally, the time and energy required to copy them
from off-chip DRAM is eliminated. The specific array architectures, weight cells, and even details of operation
are still being investigated and vary considerably across implementations. One of the key design decisions is whether
to support on-chip training (\cite{GoogleTPU, SYu1, GBurr1}), or whether weights are to be transferred to the chip after off-line training \cite{Strukov1}. Each approach
has advantages and disadvantages. On-chip training offers the most local customizability, as well as a near-complete
suppression of the impact of process variability (since the training is done for the specific hardware). The key disadvantages
are the significantly increased complexity of the peripheral circuitry needed to support on-chip backpropagation, as well
as the much higher required precision of the weights (to handle the many small increments in the weights during the iterative
training process). Implementations with off-chip training are more straightforward to program, but must contend with process
variability impacting weight values and imperfect programming \cite{Strukov1}. It should be noted that off-chip training 
with analog or multi-bit digital weights which are implemented as multiple states of a single device suffers from similar
programming complexity as on-chip programming. Due to the non-linearity of the response of all standard NVMs, the number
or duration of programming pulses to achieve a desired weight depends on the current state of the weight (\cite{SYu1, Strukov1}). 
A sense-program
iteration is therefore required to achieve the desired weight, even though the value for the target weight was computed off-line.

In this paper, the assumption is made that near-term applications for neuromorphic accelerators on mobile SoCs will not benefit from on-chip training,
which is instead relegated to the cloud. The focus is on improved inference performance and power reduction. The cross-bar array and non-volatile weight cells are designed for programming simplicity and robustness w.r.t. process and programmation variability, enabling effective transfer of off-line weights. No iterative programming is required. Additionally, all training is assumed to be centralized (with possibly crowd-sourced data).

In order to achieve reasonable immunity to process variability and programmation errors, a multi-bit digital representation
of weights is used. It has been demonstrated (in this work and
elsewhere) that high weight precision is not required for inference-only applications.

\begin{figure}[!ht]
\centering
\includegraphics[width=3.0in]{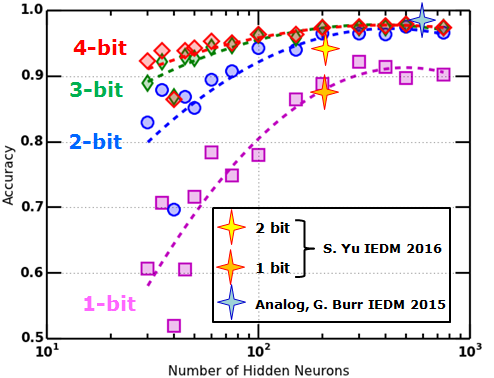}
\caption{The weight-precision/layer size tradeoff in DNN test accuracy is illustrated, using the MNIST benchmark.
A single hidden layer of varying size is simulated with varying weight precision. The weights are
obtained using standard training in software, then quantized for test accuracy evaluation. The quantization
range is optimized separately for each point; results with the optimized quantization window are shown.}
\label{weight-precision}
\end{figure}

\noindent As seen in Fig. \ref{weight-precision}, even 2-bit weights achieve near-analog levels of accuracy
with reasonably-sized, fully connected layers on a simple benchmark such as MNIST. Additionally, it is also possible to re-train
a network to use larger layers if additional accuracy is required; choosing 2-bit weights therefore does not
impose an overall inference accuracy constraint. Finally, it should be noted that the results of Fig. \ref{weight-precision}
are obtained by straightforward quantization of software-trained network (with the additional step of optimizing the
quantization window). If the network is trained with the assumption of quantized weights, even better accuracy can be obtained. 

\FloatBarrier
\section{Architecture}
\begin{figure}[!ht]
\centering
\includegraphics[width=3.5in]{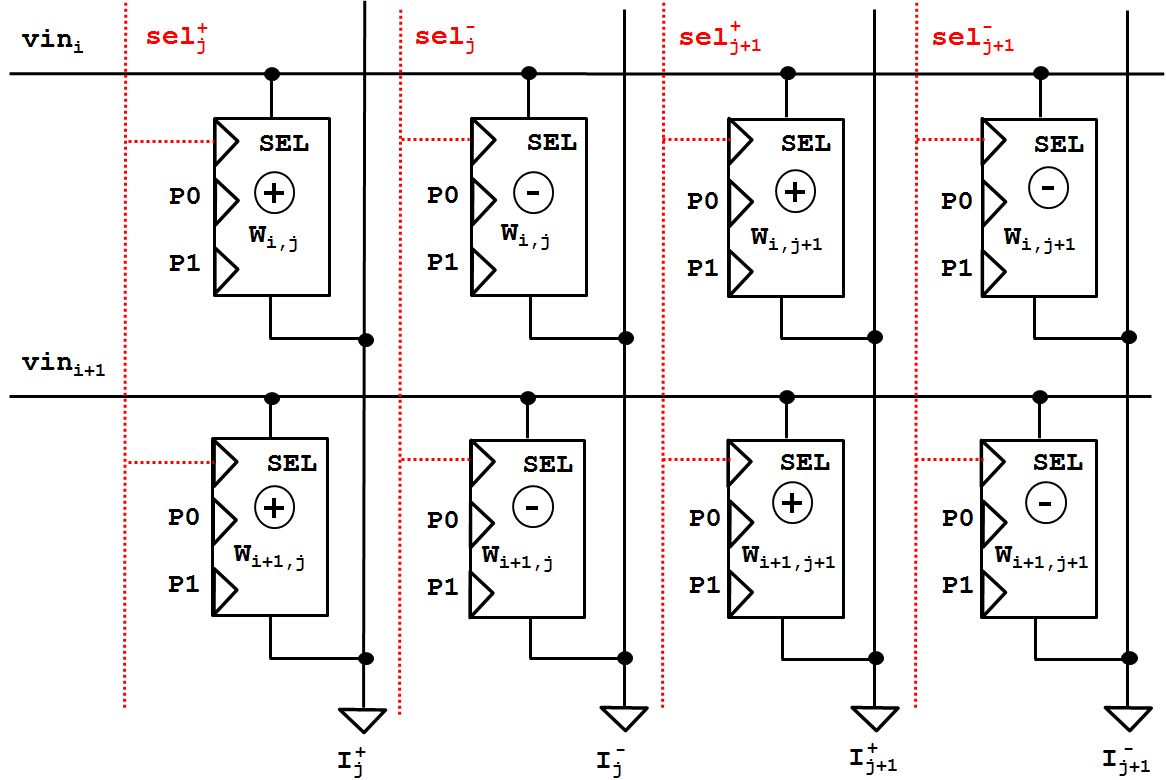}
\caption{The cross-bar array utilized in this work is illustrated. The array consists of two sets
of weights: one each for positive and negative conductance contributions. An additional feature of the array
is the set of dedicated program lines, used only during the programming events. Two-bit weights are shown, as indicated
by the two programming inputs P0 and P1. Note that each sign contribution to the weight has two bits.}
\label{array}
\end{figure}

The array architecture is a slightly modified resistive cross-bar array, as shown in Fig. \ref{array}.
The standard approach of using two weights to represent positive and negative conductances is used. The modification
of the standard approach arises only in the use of dedicated program lines; one for each bit and for each row of
weights. The program lines are shared across the entire row of weights; selecting individual weights to program
is accomplished using the select lines. In inference mode, the program lines are grounded, and the weights 
behave like two-terminal devices, forming a cross-bar between the signal input and output lines.

\begin{figure}[!ht]
\centering
\includegraphics[width=3.0in]{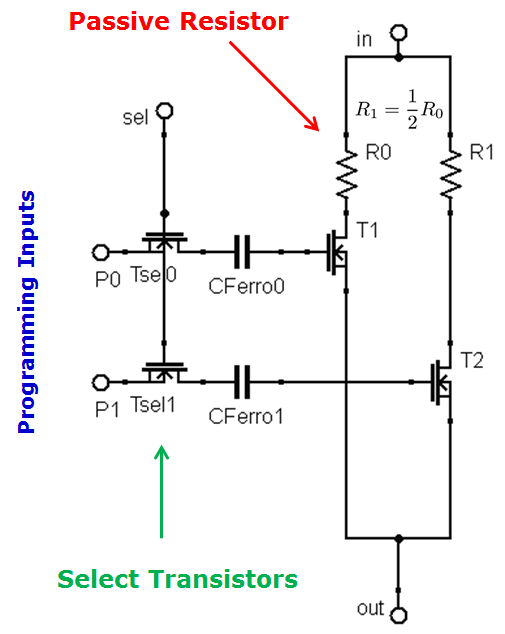}
\caption{The schematic of a two-bit weight cell is illustrated. The resistance is provided by 
passive resistors, while the NVM transistors enable or disable the individual branches. A select
transistor is provided for each bit. Additional bits require additional parallel branches
of resistor-transistor combinations. During inference, the program inputs are grounded
and the weight cell acts as a two-terminal device. The values of the passive resistors are
chosen to provide a binary ladder of overall weight conductance.}
\label{two-bit}
\end{figure}

\begin{figure}[!ht]
\centering
\includegraphics[width=2.0in]{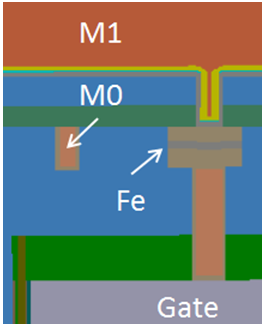}
\caption{The cross-section of an emulated FeFET structure is illustrated. The underlying FETs
and gate are at the bottom, with the ferroelectric layer and surrounding metal electrodes placed
at the M0 level.}
\label{fefet}
\end{figure}

The resistive weight
cells are composed of a parallel combination of passive resistors, with each passive resistor in series
with a gating transistor (as illustrated in Fig. \ref{two-bit}). 
The gating transistor could be a FeFET or a Flash transistor, or other FET-based NVM. 
In this work, we describe an architecture using a FeFET as the gating transistor, with the FeFET composed
of a standard FET with a ferroelectric capacitor (FeCap) in the BEOL layers.
The FeFET gate and BEOL physical structure is shown in Fig. \ref{fefet}. 
Electrically, this is similar to the integrated FeFET
structure of \cite{fefet1}.

The resistance values of the passive resistors for the weight cell are chosen as $R_0,\ \frac{1}{2}R_0,\ 
\frac{1}{4}R_0$ etc. Conversely, the conductance of each branch (neglecting the influence of the transistors)
is given by $G_0,\ 2G_0,\ 4G_0$ etc. Denoting the logic state of each transistor (indexed by $i$) as $b_i$, the
overall conductance of the cell is given as:
\begin{equation}
G_{tot} = G_0 \sum_{i=0}^{n-1}b_i 2^i
\label{conductance}
\end{equation}
\noindent where $n$ is the number of bits in the cell ($n=2$ in the example of Fig. \ref{two-bit}). The state
of the transistors is set by program and erase cycles (described in section \ref{progerase}), with the SEL
input active. During inference, the program inputs ($P_i$) are grounded, and the potential on the gate of the 
transistors is determined by the polarization state of the ferroelectric capacitors. The programming is assumed
to result in strongly-ON or strongly-OFF transistors (i.e. strongly negative or strongly positive V$_t$, respectively),
resulting in branch conductances that are (ideally) independent of the properties of the transistors or the precise
programming of the FeCaps.

\FloatBarrier	
\section{Ferroelectric Capacitor: Properties and Modeling}
\label{FeCap}

\begin{figure}[!ht]
\centering
\includegraphics[width=3.0in]{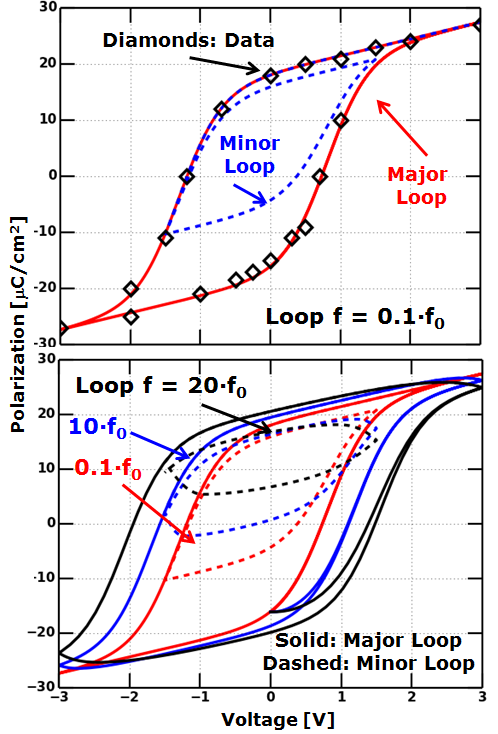}
\caption{The simulated polarization hysteresis of a HfZrO$_2$ FeCap is shown 
(measured data from \cite{ferro1}). The top
figure illustrates the calibration to data on the outer (saturation) hysteresis loop, 
as well as the modeled behavior of an example minor loop (the minor loop is achieved by moving
from positive saturation to a negative voltage equal to one-half of the max voltage, then
swinging to a positive half-max voltage). The bottom figure illustrates the modeled frequency
response. When the loop frequency is much longer than the ferroelectric domain response
frequency $f_0$, the behavior of the FeCap is quasistatic. For increased loop frequencies,
the hysteresis loop becomes quite distorted, with an apparent increase in coercive voltage
and a flattening of the minor loops. }
\label{hysteresis}
\end{figure}

The FeCaps shown in Fig. \ref{two-bit} are the enablers for the non-volatile operation of the weight cell.
The desirable property of the FeCaps is the hysteretic behavior of the polarization, as illustrated in
Fig. \ref{hysteresis} (using measured data from \cite{ferro1}).

\noindent In this work, the hysteretic behavior of FeCaps is modeled using a Preisach-based
\cite{Preisach}, 
turning-point model with explicit internal polarization dynamics. The quasi-static 
ferroelectric polarization
is described next. The "raw" response function is given by:
\begin{equation}
F^{\mp}(V_{int})= \theta^\pm \cdot tanh \bigg( \frac{V_{int} \pm V_c^\pm}{V_{sc}^\pm} \bigg)
\label{raw}
\end{equation}

\noindent where $V_{int}$ is the voltage representing the internal state of the FeCap (related
to the applied voltage, as described next), $V_c^\pm$ and $V_{sc}^\pm$ model parameters describing
the coercive voltages and the voltage scales, respectively. Likewise, $\theta^\pm$ is a model parameter
which sets the polarization strength in each state. Each quantity in Eqn. \ref{raw}
has a "plus" and "minus" label, depending on whether the capacitor last experienced an increase
or decrease in applied voltage (respectively). This is referred to as the "state" of the FeCap.
The actual ferroelectric polarization $P_{FE}$ is computed using Eqn. \ref{Polarization}:

\begin{equation}
P_{FE}(V_{int}) = \bigg( F^{\pm}(V_{int})-F^+_j \bigg) \cdot 
\bigg[ \frac{P_j-P_i}{F^+_j-F^-_i} \bigg] + P_j
\label{Polarization}
\end{equation}

\noindent where the indices $i$, $j$ denote the currently active turning points, with $V_j > V_i$.
Likewise, the quantities $F^+_j$, $F^-_i$, $P^j$, and $P^i$ are evaluated at the current active
pair of turning points. Eqn. \ref{Polarization} simply provides scaling and shifting of the raw response
of Eqn. \ref{raw} to insure that the polarization curve passes through the active pair of turning points
(this is necessary to provide reasonable minor-loop behavior). 
As is standard for turning-point models, turning points themselves are created and destroyed
dynamically as the internal voltage of the FeCap switches. The ferroelectric response is thus modeled
as a quasi-static function of the "internal" state of the FeCap (described by $V_{int}$). The overall
behavior of the FeCap is therefore dependent on the dynamics of $V_{int}$. In this work, $V_{int}$ is
governed by a second-order delay of the applied voltage across the FeCap. Specifically, the differential
equation relating the "internal" voltage $V_{int}$ and the applied voltage $V_{app}$ is given as:

\begin{equation}
\ddot{V}_{int} 
+  2 \gamma \omega_0 \dot{V}_{int} 
+ \omega_0^2 V_{int}
= \omega_0^2 V_{app}
\label{dynamics}
\end{equation}

\noindent where the natural frequency $\omega_0$ and the damping ratio $\gamma$ are calibration parameters
of the model. The "memory" property of the FeCap is handled by the model though the turning point
history and the active state. Both are changed in a discrete manner, when the temporal derivative of
$V_{int}$ changes sign. This ensures that the state itself is not experiencing the second-order dynamics
described in Eqn. \ref{dynamics}; the dynamics merely provide a delay for an abrupt switching of the state.
It should be noted that the second-order delay model of Eqn. \ref{dynamics} is purely empirical. It is merely used
in this work to provide a delayed response for teh switching of ferroelectric domains. 
No attempt is made to provide a detailed frequency-dependence
calibration to a specific material. It has been shown in the literature that the frequency response of
ferroelectric materials covers a very wide range; from a very fast responses in the 10s of ns \cite{ferro2},
to a much slower 10s of $\mu$s \cite{ferro3}. For the purposes of this work, it is assumed that the ferroelectric material
has a switching frequency comparable to the fast material of \cite{ferro2}. For materials where this is not the case,
the programming times discussed in \ref{progerase}
must be adjusted to accommodate the slower frequency response. Finally, the total
charge of the capacitor is computed as:
\begin{equation}
Q_{tot}(V_{int}, V_{app}) = \big(P(V_{int}) + C_{lin} V_{app} \big) \cdot A
\label{Charge}
\end{equation}

\noindent where $C_{lin}$ is the non-ferroelectric capacitance, and $A$ is the total capacitor area. 
The non-ferroelectric response is modeled as being driven by the instantaneous applied voltage $V_{app}$,
since the non-ferroelectric response is assumed to be much faster than any modulation of the applied voltage.
For the purpose of this work, the complete model is implemented in Verilog-A and used within Synopsys HSPICE.

\FloatBarrier
\section{Program and Erase}
\label{progerase}

\begin{figure}[!ht]
\centering
\includegraphics[width=3.0in]{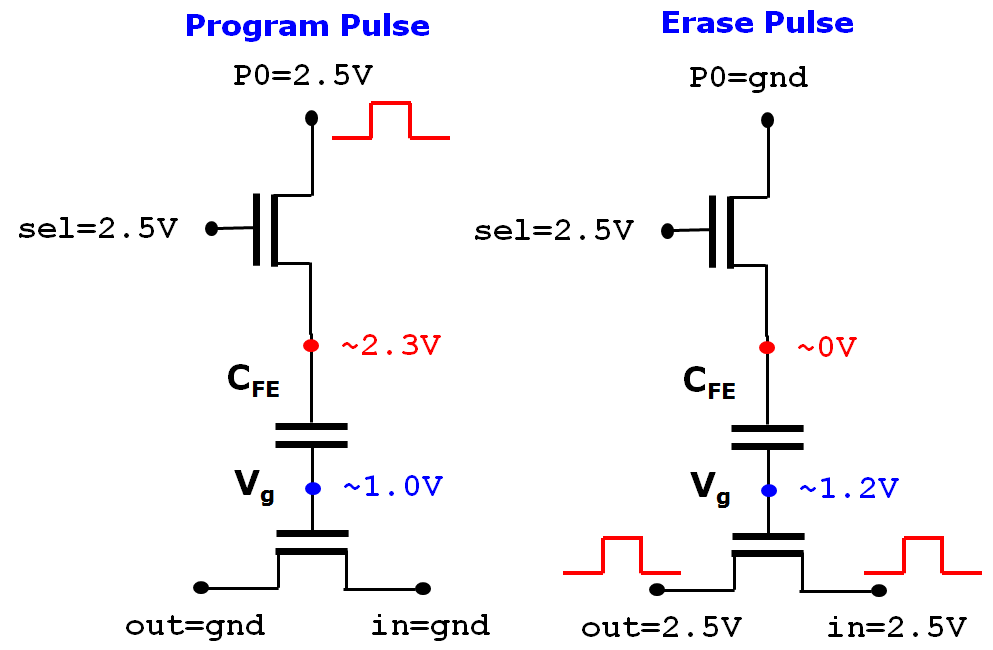}
\caption{The "Program" and "Erase" schemes for the FeCaps are illustrated. In both cases,
a suitably large voltage is established across the FeCap, with opposite polarities in the
two conditions. Due to capacitive voltage division across the FeCap and the FET (and to
a lesser extent the select transistor as well), the voltage across the FeCap is considerably
smaller than that of the applied pulse. Given the relatively large coercive voltages available
in current technologies, the programming is necessarily done at moderately high voltages. }
\label{prog_conditions}
\end{figure}

In order to assign the weights of the neural network, the FeCaps must be programmed
to appropriate states. This can be performed by applying voltage pulses to either the 
program terminals or the in/out terminals of the weight. Program and erase
schemes for the FeCaps are illustrated in Fig. \ref{prog_conditions}. In this work,
a programming scheme which results in a positive after-pulse voltage on the gate node of the underlying
FET is termed "Program," whereas a scheme which results in a negative 
after-pulse gate voltage is termed "Erase." 
In the example array of Fig. \ref{array},  a select transistor is needed in order to enable programming, 
since only a single set of program
interconnect lines can be provided for each row of the crossbar array.  
Since the weights of a given row share program lines,
the column which is being programmed at any given time is selected by activating the appropriate select line.

\begin{figure}[!ht]
\centering
\includegraphics[width=3.3in]{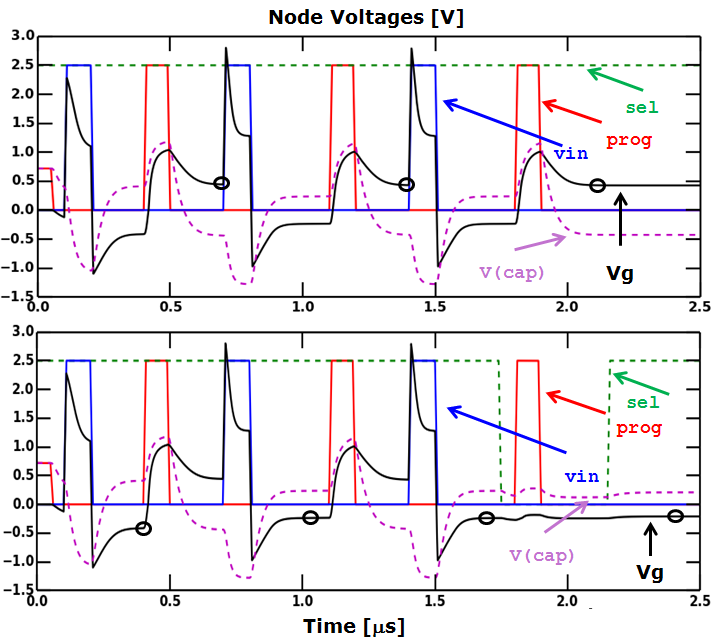}
\caption{The simulated cycle of erase and program events for a single bit.
After the initial erase and program event, the nodal voltages exhibit
steady-state behavior. The final voltage on the gate of the FET indicates
the programming strength of the bit. The program event is shown in the top figure,
the erase event is shown in the bottom figure. The non-quasistatic behavior
is almost entirely due to the delaying dynamics of the FeCap; the RC time constants
are much shorter than the clock period. }
\label{combined_erase_prog_time}
\end{figure}

\begin{figure}[!ht]
\centering
\includegraphics[width=3.5in]{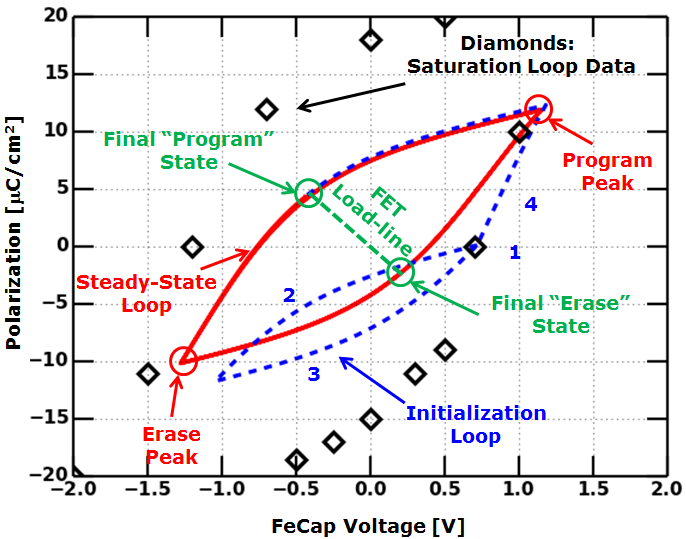}
\caption{The simulated polarization hysteresis of a set of program and erase
cycles is shown. The initialization loop is shown as a blue dashed line, while
the steady-state loop is indicated with a solid red line. The saturation loop data
(to which the model is calibrated) is shown with black symbols. Note that the 
steady-state loop is much smaller than the saturation loop. }
\label{erase_prog_erase_hyst}
\end{figure}

An additional constraint is the fact that the individual FeFETs of a given weight share
common in/out terminals. It is therefore not possible to erase them individually, using the
scheme of Fig. \ref{prog_conditions}. One possible programming strategy is therefore to 
initially fully erase an array, with all FeCaps set to the erased state, followed by individual
programming of FeCaps which need to be in the programmed states. This is somewhat akin to "flash"
erase commonly used for various Flash technologies. Both erase and program are expected to take
place infrequently: only when the neural network is initially mapped onto the SoC. During inference
operations (which are expected to be frequent), no erase or program operations take place. 
Figures \ref{combined_erase_prog_time} (top and bottom, respectively) illustrate program and
erase events in more detail. In Fig. \ref{combined_erase_prog_time}(top) a single bit is cycled several
times by alternating erase and program pulses and ultimately left in a programmed state.
The cycling is performed to ensure that the state of the bit is independent of the initial
condition. 
The P-V trajectory for the erase-program cycle is illustrated in Fig. \ref{erase_prog_erase_hyst}. 
Simulation begins with zero charge on the FeCap, in state 1. After the initial erase (arc 2), 
the first program pulse takes the FeCap to the highest achievable positive voltage, along
arcs 3 and 4 of Fig. \ref{erase_prog_erase_hyst}. This corresponds to the peak voltage point under the 
program pulse in Fig. \ref{combined_erase_prog_time}. After the programming
pulse ends (as $V_{app}$ is reduced to zero), 
the FeCap P-V trajectory continues along the steady-state loop (Fig. \ref{erase_prog_erase_hyst})
until it intersects the FET $C_{gg}$ load-line (the FET is now in strong inversion and has a nearly
constant capacitance). The FET $C_{gg}$ load line
is simply established from the equation for the voltage drop across the FeFET stack:
\begin{equation}
V_{app} = V_{cap} + V_g
\label{Kirchhoff1}
\end{equation}
\noindent After the program pulse is completed, Eqn. \ref{Kirchhoff1} reads simply $V_{cap}=-V_{g}$, 
and with the condition that $Q_{cap}=Q_g$, the loadline can be established as
\begin{equation}
P_{cap}  \approx -\frac{C_{gg}}{A_{cap}} \cdot V_{cap}
\label{loadline}
\end{equation}
\noindent The equation for the loadline is approximate, since the FET gate capacitance is not constant.
The final voltage across the FeCap after a programming even is negative (approximately -0.5V). 
Correspondingly, the voltage on the gate node of the FET is approximately
positive 0.5V, putting the FET into a (relatively) weak ON-state. The is the "Programmed" condition.
The application of the erase pulse (achieved by simultaneously pulsing the in/out terminals while grounding
the program terminal) makes the FeCap voltage more negative, continuing the P-V trajectory from the 
program point,
all the way to the most negative voltage on the steady-state loop. 
After the erase pulse is completed, the P-V trajectory continues
until it intersects the FET $C_g$ load-line at a slightly positive voltage. The slightly
positive voltage of the FeCap with a grounded program line implies a slightly negative voltage on
the gate of the FET (simulations indicate approximately -0.25V). Given that a typical Vt of the FET is
approximately 350 mV, the erase cycle has left it in a strongly OFF condition. This is the "Erased" condition.

While Fig. \ref{combined_erase_prog_time}(top) left the FeCap in a programmed state, 
Fig. \ref{combined_erase_prog_time}(bottom)
illustrates how the erase (or rather, the "non-program") is accomplished instead. 
Since all FeCaps are initially erased, the select transistor
is used to make sure that program events don't change the state of FeCaps which must remain erased.
The \textbf{sel} input is set to \textbf{gnd} during the last program pulse of \ref{combined_erase_prog_time}, preventing
programming from taking place. The final voltage on the gate of the FET is equal to that of the initial
erased state. 
It should be noted from Fig. \ref{erase_prog_erase_hyst} that due to voltage division between the FeCap and the FET,
the FeCaps in this example
operate on a minor loop which is much smaller than the outer saturation loop. This indicates that much stronger
programming is theoretically possible. In order to reach the saturation loop and correspondingly stronger
ON-states of the FET, significantly higher programming voltages are required. Furthermore, the use 
of an nFET for the select device implies that the highest voltage across the FeFET is the lesser of
$V_{select}-V_t$ and $V_{program}$. High programming voltages therefore imply high $V_{gs}$ values for
the select device, suggesting that a high voltage design is required.

\begin{figure}[!ht]
\centering
\includegraphics[width=3.5in]{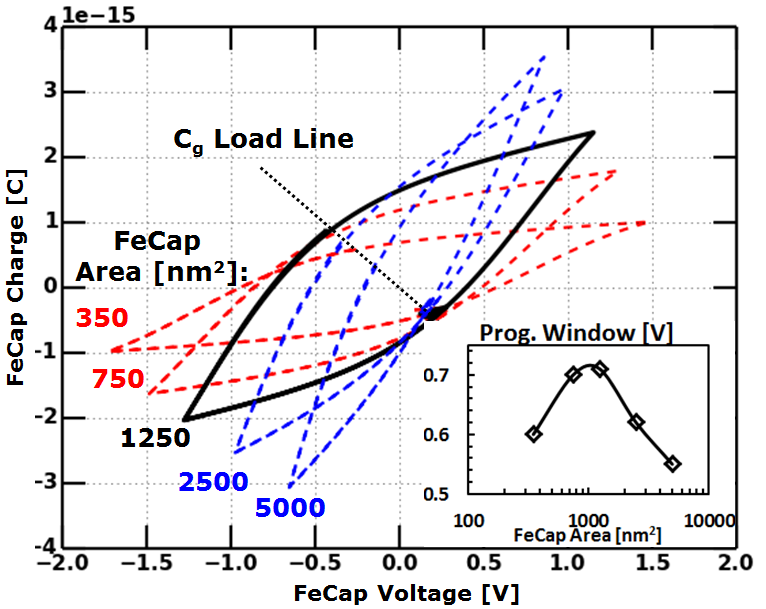}
\caption{The effect of FeCap area on the program and erase conditions is illustrated.
The behavior of  $\Delta V_{prog}$ is clearly non-monotonic w.r.t. FeCap area. Very small
FeCaps have large peak programming voltages, but reduced $\Delta V_{prog}$ due to an overall
reduction in charge (small area). Very large
FeCaps have small programming voltages since most of the applied voltage is dropped across
the FET gate, and likewise result in reduced$\Delta V_{prog}$.  }
\label{area}
\end{figure}

In this work, since the FeCaps are implemented in the BEOL levels, the area of the FeCaps is tunable and there is 
no layout area penalty associated with FeCap size.
It is therefore
possible to choose the FeCap area which maximizes the $\Delta V_{prog}=V_{prog}-V_{erase}$ voltage
window. The behavior of the program and erase loop as a function of FeCap area is
illustrated in Fig. \ref{area}. Increasing the FeCap area results in a smaller voltage
drop across the FeCap during programming. This constricts the size of the minor loop
for the program/erase cycle, with lower end voltages and lower polarizations. This is
seen in Fig. \ref{area} as a tightening of the polarization loop with increasing area.
At the same time, the increased area increases the total charge, resulting in higher
peak charge values on the FeCap and FET gate (in spite of the reduced charge/unit area).
This is reflected in Fig. \ref{area} as higher peak charge values at the ends of the 
polarization loops. The final program and erase voltages are obtained from the intersection
of each polarization loop with the zero-$V_{app}$ $C_{gg}$ load line of the FET. 

It can be seen
that the locus of the intersection varies non-monotonically with FeCap area (especially for the
program portion of the loop), due to the competing
effects of reducing polarization and increasing area. The maximum in the $\Delta V_{prog}=V_{prog}-V_{erase}$ voltage
window is obtained for an FeCap area of 1250 nm$^2$ (given the particulars of the underlying FETs and
the FeCap hysteresis curve). Significantly larger or smaller areas result in degraded programming
performance.

\FloatBarrier
\section{Passive Resistor}
\label{sec:resistor}
An important element of the multi-bit weight cell of Fig. \ref{two-bit} is the passive resistor. 
The resistance value of the resistor could be chosen so that when the FeFET in a branch
of the cell is in the ON-state, the resistance of the passive resistor is much larger
than that of the FET. If this is the case, then the non-linear behavior of the FET
will have a negligible effect on the overall conductance of the cell. Given that the 
resistance of the FET is expected to be in the range of a few $k \Omega$, the passive
resistor should have a resistance in the range of $30 k\Omega$ to $100 k\Omega$. Larger
values of resistance result in better overall linearity, but also limit the current.
With the stated resistance range, currents supplying the summing amplifiers are in the
reasonable range of a few to a few tens of $\mu A$. There are many options for implementing
the passive resistor, with varying tradeoffs on size, variability and linearity. One possible
approach is to simply use the channel of a FET as the body of the passive resistor.



\begin{figure}[!ht]
\centering
\includegraphics[width=3.0in]{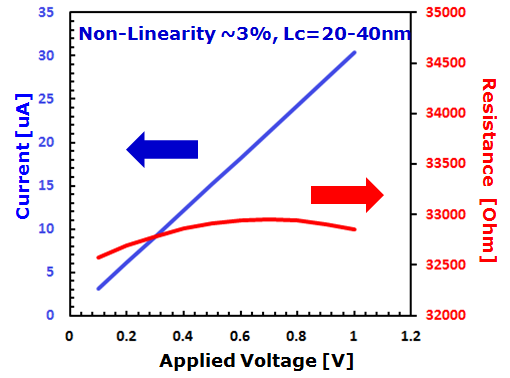}
\caption{The resistor current and resistance values are shown. The target
resistance values of a few tens of $k \Omega$ appear to be achievable,
and the value of the resistance is very nearly constant over a wide
range of applied voltages. }
\label{resistor_current}
\end{figure}

Simulations of such resistors (Fig. \ref{resistor_current}) indicate that the desired
resistance level can be achieved with reasonably good linearity.

\begin{figure}[!ht]
\centering
\includegraphics[width=3.0in]{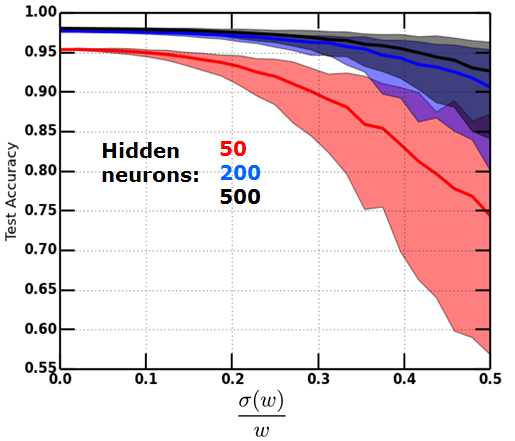}
\caption{The effect of uncorrelated resistor noise due to RDF-induced
variability is evaluated by simulation. The expected regime of noise
for the passive resistors described in this work is $<$ 10\%. With reasonable
hidden-layer sizes, the effect of this level of noise is negligible. }
\label{noise}
\end{figure}

A concern with doped resistors is the
RDF-induced variability. Unlike a FET with a doped channel, the question is not
one of Vt-variability, but simply that of variable "bulk" conductance due to
the random number of dopants in any one resistor. As such, the variability
of conductance (assuming Poisson-distributed number of dopants per channel) can be
expressed as:
\begin{equation}
\frac{\sigma(G)}{G} = \frac{1}{\sqrt{N}}
\label{RDF}
\end{equation}

\noindent where $G$ is the conductance of the resistor body, $\sigma(G)$ is the 
RDF-induced standard deviation of the conductance, and $N$ is the total (expected)
number of atoms in the resistor body. Given reasonable resistor dimensions, lengths, and
target doping values, the expected number of atoms is \textasciitilde 100, resulting in a relative
standard deviation of \textasciitilde 10\%. While this may seem like a large uncertainty in the
conductance (and therefore in the neural net weight), it is of relatively little
consequence in typical fully-connected layers. This is illustrated in Fig. \ref{noise},
where the accuracy of a neural net is evaluated with various levels of uncorrelated noise.

A single hidden-layer network with various sizes of the hidden layer is trained for MNIST.
The weights are quantized to two-bit precision and perturbed using a Poisson distribution,
as indicated by Eqn. \ref{RDF}. A set of MC simulations is then performed with various levels
of perturbation. As can be seen from Fig. \ref{noise}, there is no significant loss of 
classification accuracy until the relative weight perturbation is on the order of 30\%.
This is not surprising for for fully-connected architectures, such as the one used. The 
large number of inputs to any one neuron induce partial noise cancellation, with the
expected standard deviation of the total input to the neuron scaling by $1/\sqrt(N_{inp})$, 
$N_{inp}$ being the number of inputs. The larger the network layers then, the less susceptibility
to uncorrelated weight noise there is. Given the expected level of $<$ 10\%, the RDF of the
resistors appears to be satisfactory.

\FloatBarrier
\section{Transistor}

\begin{figure}[!ht]
\centering
\includegraphics[width=3.5in]{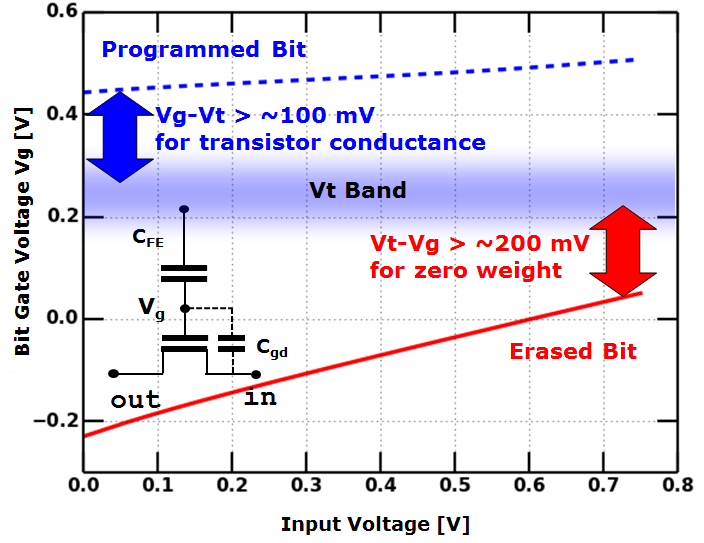}
\caption{The allowed Vt-band of the underlying FETs is illustrated, along with
the behavior of Vg with applied input voltage. The Vt
must be low enough so that the programmed Vg exceeds it by at least 100 mV, enabling
sufficient gate overdrive to keep the conductance of the transistor much higher than
that of the passive resistor. Likewise, the Vt must also be high enough so that 
the Vg of the erased bits is always at least approximately 200 mV below it to make sure that
the FET never turns on significantly. Satisfying both criteria simultaneously
results in a narrow band of allowable Vt values. The Vg of the FET varies with the applied
input voltage due to capacitive coupling of the drain and gate (both parasitic and intrinsic).  }
\label{vt}
\end{figure}

An important parameter of the underlying FET of the FeFET structure is the threshold voltage. 
There are typically two or three Vt levels provided, but the FeFET requirements place restrictions
on the choice. This is illustrated in Fig. \ref{vt}.

\noindent As seen in Fig. \ref{vt}, the application of the input voltage to the weight cell causes
a slight shift in the Vg node voltage. This is a result of the capacitive coupling of the FET
drain and gate, both due to the parasitic and intrinsic coupled charge. The effect is particularly pronounced
on erased bits, since very little current flows in the erased branch, and essentially the entire input voltage
is applied to the drain of the FET. The IR drop across the resistor lessens the impact in branches with 
programmed bits. Fig. \ref{vt} illustrates a possible band of Vt values for the FET. For all applied
input voltages, the erased transistor has a gate voltage significantly below the FET Vt. This ensures that
the erased transistor remains turned off (the extent to which it is turned off is a technology parameter that sets
the max ON/OFF ratio of the weights). The Vt is also low enough so that the programmed transistor has sufficient
gate overdrive to keep the transistor conductance much higher than that of the passive resistor. Thus, the choice
of the Vt falls into a somewhat narrow band, and induces a tradeoff between weight accuracy and linearity on one hand
(Vt must be low), and low conductance of the zero weights (Vt must be high).
Vt values in the 200-300 mV range needed for the implementation shown in this paper are typically readily available in
modern CMOS technologies.

\FloatBarrier
\section{Results}

\begin{figure}[!ht]
\centering
\includegraphics[width=3.5in]{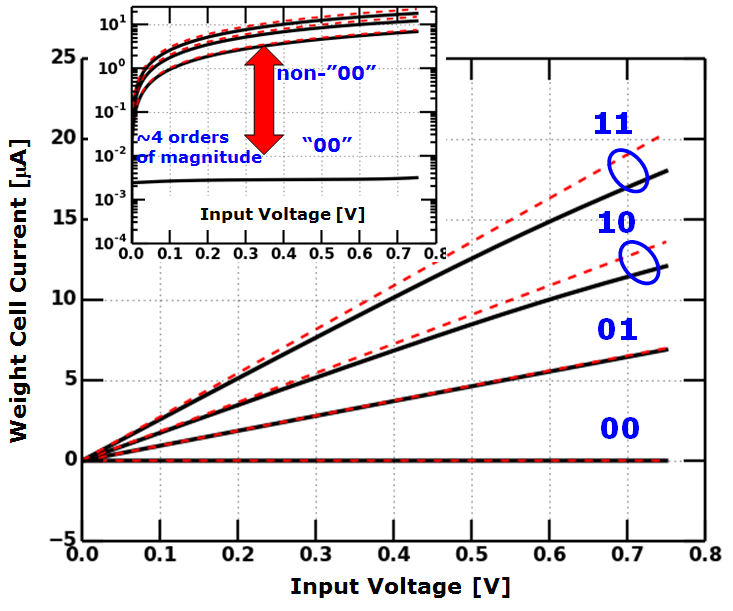}
\caption{The I-V relationship for a two-bit weight cell is illustrated for
all four programming combinations. The simulated current is shown in black, while
the ideal currents (as if the weight cells were ideal resistors) are shown as
dashed red lines. The inset shows the same currents on a logarithmic scale, illustrating
the large dynamic range of currents available. }
\label{currents}
\end{figure}

Having programmed the individual FeFETs in the array, inference can be performed simply by 
applying input voltages (fixed levels or pulses) to the array. The program lines are all grounded
during inference. In this mode, each weigh cell acts as a two-terminal device. For the case of
the two-bit cell, the I-V characteristics are illustrated in Fig. \ref{currents}.

\begin{figure}[!ht]
\centering
\includegraphics[width=3.5in]{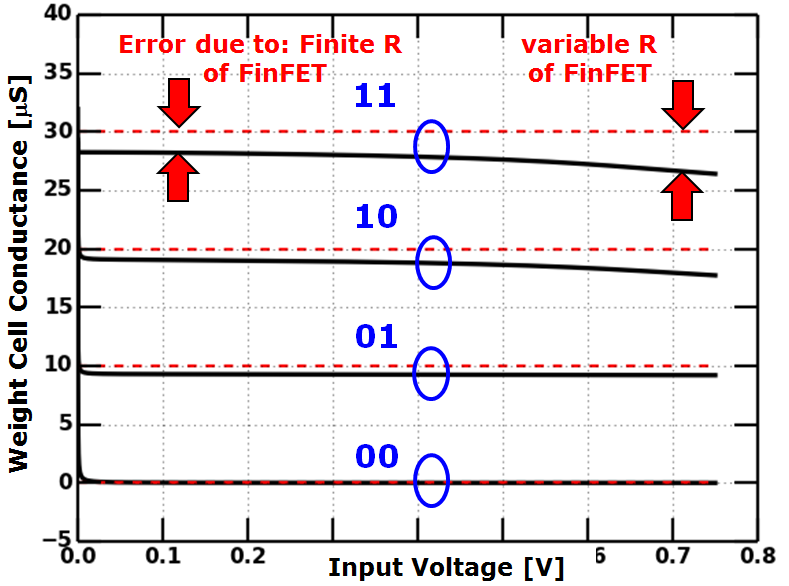}
\caption{The conductances of the four states of the two-bit weight cell are shown.
Solid black lines are simulated conductances, dashed red lines represent ideal values.
The error at low input voltage is attributable to the finite resistance of the FET
in linear mode. At high voltages, the error increases due to the larger FET
resistance in saturation. }
\label{weights}
\end{figure}

The I-V characteristics of Fig. \ref{currents} show good linearity over a wide range of 
input voltages, with appreciable deviation seen only for the highest current conditions.
The dynamic range of ON/OFF currents is roughly four orders of magnitude, considerably larger
than for memristive devices. This makes the multi-bit weight cell suitable for arrays
with large numbers of inputs. The accuracy of the various ON-states is further illustrated
in Fig. \ref{weights}.

\begin{figure}[!ht]
\centering
\includegraphics[width=3.5in]{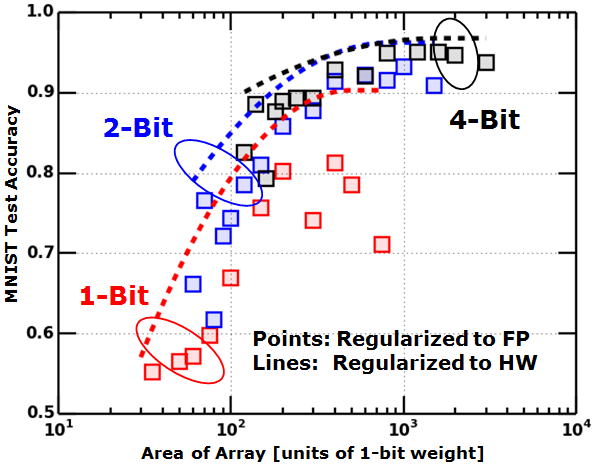}
\caption{The test accuracy on the MNIST benchmark is computed with
weight models describing the FeFET-based weight cells. Weights are
quantized to 1, 2, and 4-bit precision. The quantized weights are non-linear
as described by Fig. \ref{weights}. All activations are binarized. The accuracy
is plotted vs. the total array area (number of cells times the cell area).}
\label{final}
\end{figure}

\noindent It is evident from Fig. \ref{weights} that the absolute accuracy of the weights
is quite good when the weights are small, but  the weights become increasingly less accurate
as the nominal weight value increases. This is simply a result of the finite (and non-linear)
resistance of the FET increasing the total series resistance. The effect is negligible for the
small weights since the resistance of the passive resistor in those cases is large.  
Similarly, the accuracy is best for small values of the input signal but shows increasing error
with the signal voltage. This is due to the transition of the FET from the linear to the
saturation regime as the $V_{ds}$ across the FET increases (particularly sensitive for
large weight branches where the resistance of the passive resistor is small). Given the
expected deviations of the hardware weights from ideality, it is necessary to simulate
the accuracy of the overall neural net. The standard simple MNIST benchmark is used, 
as shown in Fig. \ref{final}. In addition to the quantized weights as described in this work,
the neuronal activations are considered to be binarized. While this is certainly not necessary,
binarization greatly simplifies the problem of transferring signals from one layer to the next,
eliminating the need for ADC/DAC conversion.

As can be seen in Fig. \ref{final}, the networks with non-ideal weights and
binarized activations do indeed suffer some accuracy loss, as compared
to ideal multi-bit weights (Fig. \ref{weight-precision}). The degree of accuracy
loss depends on the regularization procedure, however. Points in Fig. \ref{final}
are obtained by direct quantization of weights from software-based training.
Dashed lines are obtained using weights which were trained
using hardware-aware regularization. While both sets of data use regularization,
the inference on the validation set for the second group was performed using
the full hardware model, including weight quantization, weight non-idealities, 
and binarization of activations. The algorithm is summarized in Alg. \ref{algo}. 
The  validation accuracy thus obtained is a much better
estimate of the final test accuracy (which also uses non-ideal weights and activations), yielding
much better test accuracies overall. This is particularly evident in cases where the
neural network is over-provisioned (large arrays), and significant over-fitting occurs
if the network is not properly regularized. With hardware-aware regularization, it can
be seen that the test accuracy does not degrade for large arrays, and test accuracies
are significantly better overall. Even better results should be obtainable if the
training procedure itself is hardware-aware \cite{xnornet}.

\begin{algorithm}
    \SetKwInOut{Input}{Input}
    \SetKwInOut{Output}{Output}

    \underline{Regularization Algorithm} $(train, validate, \lambda)$\;
    \Input{Training set $train$, validation set $validate$, $\lambda_{min}$, $\lambda_{max}$}
    \Output{$w_{optim}$}
    
		\For{$\lambda_i=\lambda_{min}$ to $\lambda_{max}$}{
			\textbf{Set} $w_i$ = $Train(\lambda_i,\ train)$ \\
			\For{$qw_j=qw_{min}$ to $qw_{max}$}{
					\textbf{Set} $w_{ij}^{HW}$ = $HWModel(w_i,\ qw_j,\ validate)$ \\
				}
				\textbf{Set} $cost_{ij}$ = $Validate(w_{ij}^{HW})$ \\
			}  
		\Return{$w^{HW}(argmin(cost_{ij}))$}
    \caption{Hardware-Aware $L^2$ Regularization}
		\label{algo}
\end{algorithm}

 \section{Summary and Conclusion}

A multi-bit weight cell for analog MAC in neuromorphic arrays was presented. The cell supports
in-memory computing based on NVM-storage
of weights, removing the need for DRAM access during inference (unlike GPU-like implementations
such as \cite{GoogleTPU}). It was shown
that for inference purposes, a small number of bits are sufficient for analog-like accuracy.
The proposed multi-bit weight cell uses an FeFET as gating element in each branch of a circuit
consisting of a parallel combination branches of passive resistors. 
The resistor weights are arranged in a binary ladder, enabling a uniform distribution of conductances
to be programmed into the cell. The FeFETs are constructed using standard FETs as the underlying FET
element, with FeCaps in the BEOL connected to FET gates. The polarization state of the FeCaps stores
the individual bits of the weight. It was suggested that passive resistors can be formed using 
FETs.
The program and erase cycles for the cell were described, and it was found
that moderately high voltages are needed for acceptable programming levels.
It was found that up to 700 mV of differential programming (difference
in FET Vt between programmed and erased states) could be achieved using standard CMOS processes and 
literature-based FeCap properties. With this level of programming, an acceptable level of weight linearity
could be achieved, along with an ON/OFF ratio for the weights in excess of four orders of magnitude.
Finally, it was shown that even with the non-ideal weights formed by the multi-bit circuits, a high level
of inference accuracy is possible (using the MNIST benchmark). This is particularly true if the regularization
is hardware-aware, i.e. cross-validation is performed using the actual weight model, with an optimized quantization window. . Failure to do so results
in somewhat sub-optimal weight transfer from software to hardware, and an associated degradation in test accuracy. 
It was also noted that the area efficiency
of the cell is independent of the number of bits used (except for the 1-bit case); at matched array area, the
2-bit and 4-bit cells produced the same level of inference accuracy. An unrecoverable loss was only observed
with 1-bit cells. 
Further improvements in weight ideality require improved FeCap properties. Increased ferroelectric polarization
(relative to the non-ferroelectric polarization) would result in increased Vt differentials between the programmed
and erased states, enabling better linearity and further improved ON/OFF ratio.



%



\ifCLASSOPTIONcaptionsoff
  \newpage
\fi



%
\FloatBarrier

\end{document}